\documentclass[letterpaper,superscriptaddress,aps,pra,nolongbibliography,twocolumn,showpacs,floatfix]{revtex4-1} % {{{
\usepackage{graphicx,helvet}
\usepackage{color}
\usepackage{bm}                      % added by TG
\usepackage{float}
\usepackage{tikz}
\usepackage{braket}
\usepackage{soul}
\usepackage{amsfonts}
\usepackage{amsmath}
\usepackage{lipsum}% http://ctan.org/pkg/lipsum
\definecolor{mygray}{gray}{0.4}
\definecolor{light-blue}{rgb}{0.8,0.85,1}
\graphicspath{{images/}}
\usepackage{amsmath}%
\usepackage{MnSymbol}%
\usepackage{wasysym}%
\usetikzlibrary{decorations.shapes}
\usepackage[draft,inline,nomargin]{fixme} \fxsetup{theme=color} %fxnote

\DeclareMathOperator{\Tr}{Tr}

\newcommand{\q}[1]{``#1''}

\newcommand{\regens}{Institut f\"ur Theoretische Physik, Universit\"at Regensburg, D-93040 Regensburg, Germany}

\begin{document}
	\title{Strong Coupling and non-Markovian Effects in the Statistical Notion of Temperature}
	\author{Camilo Moreno} \affiliation{\regens}
	\author{Juan-Diego Urbina} \affiliation{\regens}
	\email{camilo-alfonso.moreno-jaimes@stud.uni-regensburg.de}

\begin{abstract} \small
We investigate the emergence of temperature $T$ in the system-plus-reservoir paradigm starting from the fundamental microcanonical scenario at total fixed energy $E$ where, contrary to the canonical approach, $T=T(E)$ is not a control parameter but a derived auxiliary concept. As shown by Schwinger for the regime of weak coupling $\gamma$ between subsystems, $T(E)$ emerges from the saddle-point analysis leading to the ensemble equivalence up to corrections ${\cal O}(1/\sqrt{N})$ in the number of particles $N$ that defines the thermodynamic limit. By extending these ideas for finite $\gamma$, while keeping $N\to \infty$, we provide a consistent generalization of temperature $T(E,\gamma)$ in strongly coupled systems and we illustrate its main features for the specific model of Quantum Brownian Motion where it leads to consistent microcanonical thermodynamics. Interestingly, while this $T(E,\gamma)$ is a monotonically increasing function of the total energy $E$, its dependence with $\gamma$ is a purely quantum effect notably visible near the ground state energy, and for large energies differs for Markovian and non-Markovian regimes. 
\end{abstract}
\maketitle
\section{Introduction}
In the context of statistical physics there are two ways to explain how a system ${\cal A}$ acquires a property associated with the thermodynamic notion of temperature \cite{libroPathria,Libro_Greiner}. In the first approach, one considers the system as \textit{weakly} coupled with a thermal bath ${\cal B}$ that is initially in a canonical state at temperature $T$. If we wait long enough, ${\cal A}$ will equilibrate (in the sense of stationarity of macroscopic observables) and acquire itself a canonical distribution at the same $T$. Here, therefore, the idea of temperature is pre-assumed from the beginning. In the second approach, one considers instead that the global system ${\cal A}+{\cal B}$ is in a microcanonical distribution at total energy $E$ (we agree with \cite{Hänggi1,Campisi,Franzosi} that this is the conceptually foundational starting point to understand the meaning of temperature). Here ${\cal A}$ and ${\cal B}$ equilibrate due to the presence of a \textit{weak} interaction term, and the temperature will emerge as a parameter that fixes the condition of equilibrium. The temperature $T=T(E)$ is then a derived rather than a fundamental quantity. 

As it has been shown when going beyond the assumption of weak interactions, for strongly coupled ${\cal A}$ and ${\cal B}$ deviations from the standard thermodynamics emerge \cite{entropy,entropy2,strong3,strong4}, as well as problems defining local temperature \cite{effective,Hänggi2,Temp2,therm_equil}. Also, the equivalence between the microcanonical and the canonical approach does not hold, correlations between system and bath become important, and the system is non-extensive by nature \cite{Strong_termo,Nature,Petruccione}. In this context it is well known that when ${\cal A}$ is strongly coupled to a thermal bath the long time steady state of the system, contrary to the weak coupling scenario, does not take the Boltzmann form, neither in the open-quantum system approach \cite{Fleming,Chem,23}, in the global closed thermal state scenario \cite{canonical}, nor in the pure state setup \cite{popescue1,popescue2}.  

Here we will provide a consistent definition of temperature $T$ in the system-plus-bath scenario with arbitrary coupling strength $\gamma$ by starting with a global microcanonical state at energy $E$ and generalizing the saddle point analysis of ensemble equivalence pioneered by Schwinger \cite{Schwinger, libroQSFT}. 

The paper is organized as follows. In section II we review the relevant aspects of the emergence of temperature in the weak coupling scenario, then in section III we generalize this idea to the finite coupling case, and as an experimental relevant application, we will present the main features of this definition of temperature $T(E,\gamma)$ in the solvable case of Quantum Brownian Motion. We give some general conclusions in section IV and an outlook overview in the last section V. 

\section{The statistical emergence of temperature from a saddle-point condition }\label{section1}
Let us first review the weak coupling case by considering two many-body systems ${\cal A}$ and ${\cal B}$ with Hamiltonians $\hat{H}_{\cal A}$ and $\hat{H}_{\cal B}$, that in isolation have fixed energies $E_{\cal A}$, $E_{\cal B}$. When brought into \textit{weak} thermal contact allowing them to interchange energy through a small interaction term such that $E_{int}(\gamma)\ll E_{\mathcal{A}}+E_{\mathcal{B}}$, in equilibrium the resulting global state is microcanonical  
\begin{align} \label{micro1}
\hat{\rho}_{{\cal AB}}=\frac{\delta(E-\hat{H}_{{\cal AB}})}{\mathcal{G}_{{\cal AB}}(E)},
\end{align}      
with total energy $E=E_{\cal A}+E_{\cal B}+\mathcal{O}(\gamma)$. Here $\hat{H}_{{\cal AB}}= \hat{H}_{\cal A}+\hat{H}_{\cal B}+\mathcal{O}(\gamma)$ acts in $\mathcal{H}_{\cal A} \otimes \mathcal{H}_{\cal B}$ and $\mathcal{G}_{{\cal AB}}(E)=\Tr_{{\cal AB}} \delta(E-\hat{H}_{{\cal AB}})$ is the microcanonical partition function, the central quantity connecting statistics and thermodynamics through the Boltzmann equation
\begin{equation}\label{entropy}
	S(E)=K \log \mathcal{G}_{{\cal AB}}(E),
\end{equation} 
for the thermodynamic entropy $S(E)$, where $K$ is the Boltzmann constant. 

The emergence of temperature in this microcanonical, weak-limit scenario starts with writing the density of states, expanding the operator-valued Dirac delta function \cite{pie}, as  
\begin{equation}
\label{eq:GAB}
\begin{aligned}
\mathcal{G}_{{\cal AB}}(E)&=\Tr_{{\cal AB}}\,\delta(E-\hat{H}_{\cal AB})  \\ &=\frac{1}{2\pi}\int_{-\infty}^{\infty}d\tau\, e^{iE\tau}{\Tr_{{\cal AB}}\Big[e^{-i\tau \hat{H}_{{\cal AB}}}\Big]}, 
\end{aligned}
\end{equation}   
where $\tau$ is an integration variable with units of inverse energy. Defining $\Tr_{{\cal AB}}[e^{-i\tau \hat{H}_{{\cal AB}}}]=Z_{{\cal AB}}(i\tau)$, Eq.~(\ref{eq:GAB}) takes the form
  \begin{equation} \label{density}
   \mathcal{G}_{{\cal AB}}(E)= \frac{1}{2\pi}\int_{-\infty}^{\infty}d\tau \, e^{i E \tau} e^{\log Z_{{\cal AB}}(i\tau)}, 
  \end{equation}  
with an associated action $\phi$ 
\begin{equation} \label{phase}
\begin{aligned}
 \phi(E,\tau)& =iE\tau+\log\, Z_{{\cal AB}}(i\tau) \\ &=iE\tau+\log\, Z_{{\cal A}}(i\tau)+\log\, Z_{{\cal B}}(i\tau),
\end{aligned}
\end{equation}
where the decomposition $Z_{{\cal AB}}=Z_{{\cal A}}+Z_{{\cal B}}$ with $Z_{{\cal A}}=\Tr_{{\cal A}}[e^{-i\tau \hat{H}_{{\cal A}}}]$ and $Z_{{\cal B}}=\Tr_{{\cal B}}[e^{-i\tau \hat{H}_{{\cal B}}}]$ is possible because the interaction term is small enough to be neglected. Following an idea due to Schwinger \cite{Schwinger}, for a large number of degrees of freedom $N\to \infty$, the quantities $E$ and $\log Z_{{\cal AB}}$ are large (allowing rapid oscillations in the exponential) and we can solve the integral in Eq.~(\ref{density}) using the Saddle-Point-Approximation (SPA). The saddle-point condition 
\begin{equation}\label{SPA}   
 \frac{d}{d\tau}\phi(E,\tau)\Bigg|_{\tau=\tau^{*}}\overset{!}{=}0
  \end{equation} 
admits an analytical continuation over the lower half of the complex $\tau$ plane where we find the sole saddle-point $\tau^{*}=-i\beta$, with $\beta$ satisfying
\begin{equation}\label{condition}
i E+i\frac{d}{d\beta}\log Z_{{\cal A}}(\beta)\Bigg|_{\beta=i\tau^{*}}+i\frac{d}{d\beta}\log Z_{{\cal B}}(\beta)\Bigg|_{\beta=i\tau^{*}}\overset{!}{=}0.
\end{equation}

By interpreting the real solution $\beta=1/KT$ as the inverse temperature and $Z_i(\beta)=Z_i(i\tau^{*})$ as the canonical partition function, then $\bar{E}_i=-\frac{d}{d\beta}\log Z_i(\beta)$ is the mean internal energy of each subsystem, and the relation
\begin{equation}\label{equi}
E=\bar{E}_{{\cal A}}(\beta=i\tau^{*})+\bar{E}_{{\cal B}}(\beta=i\tau^{*})
\end{equation}
gives a condition on how the total energy is distributed between systems ${{\cal A}}$ and ${{\cal B}}$ when they are brought into contact. This microscopic analysis is thus used as the definition of both thermal equilibrium and of the inverse temperature that fixes this condition. 

Before we make this microscopic construction complete and see how $\beta(E)$ can indeed be interpreted as the thermodynamic temperature, we find important to complete the analysis of \cite{Schwinger} by discussing the regime of validity of the SPA as well as the behavior of the error terms in the thermodynamic limit. In order to bring Eqs.~(\ref{density},\ref{phase}) to the form required by the SPA, we consider first the situation where the total number of particles $N=N_{{\cal A}}+N_{{\cal B}}$ and energy $E$ are distributed in such a way that the ratios $\nu_{{\cal A}}=N_{{\cal A}}/N,\nu_{{\cal B}}=N_{{\cal B}}/N,u=E/N$ converge to non-zero constants when $N \to \infty$. Within this usual definition of thermodynamic limit applied to {\it both} subsystems ${\cal A},{\cal B}$, the ratios $\log\, z_{{\cal A}}=N_{{\cal A}}^{-1}\log Z_{{\cal A}},\log\, z_{{\cal B}}=N_{{\cal B}}^{-1}\log Z_{{\cal B}}$ also converge to finite values, and the phase $\phi(E,\tau)$ in Eq.~(\ref{phase}) takes the form
\begin{equation} \label{phase2}
 \phi(E,\tau)=N\left[iu\tau+\nu_{{\cal A}}\log\, z_{{\cal A}}(i\tau)+\nu_{{\cal B}}\log\, z_{{\cal B}}(i\tau)\right].
\end{equation}
Substitution of Eq.~(\ref{phase2}) brings Eq.~(\ref{density}) to the form suitable for SPA, and rigorously identifies the large parameter as $N$. From the general theory we then conclude that the error in the evaluation of $\mathcal{G}_{{\cal AB}}(E)$ in Eq.~(\ref{density}) by means of SPA is of order ${\cal O}(1 / \sqrt{N})$ and it remains bounded as long as $d^{2} \phi / d\tau^{2} \ne 0$.

With these observations in mind, let us use Eqns.~(\ref{entropy}) and (\ref{density}) with the solution established by Eq.~(\ref{condition}) to obtain
\begin{equation} \label{equiv2}
\mathcal{G}_{{\cal AB}}(E)=\mathrm{e}^{\frac{S(E)}{K}} = \frac{\mathrm{e}^{\beta E+\log Z_{{\cal AB}}(\beta)}}{\sqrt{2\pi \frac{\partial^2\log Z_{{\cal AB}}(\beta)}{\partial \beta^2}}}\left[1+\mathcal{O}(N^{-1/2})\right],
\end{equation}
which by introducing the Helmholtz free energy $F(\beta)=-\frac{1}{\beta}\, \log Z_{{\cal AB}}(\beta)$,  immediately gives the well known thermodynamic relation \cite{libroQSFT}
\begin{equation}
S(E)/K=\beta E-\beta F(\beta)\,\Longleftrightarrow F=E-TS(E).
\end{equation} 
If there exist many solutions to Eq.~(\ref{SPA}) over the imaginary $\tau$ axes, say: $\tau^{*}_1=-i\beta_1, \tau^{*}_2=-i\beta_2$ with $S_1(E) > S_2(E)$, we must choose the solution $\tau^{*}_1$  in Eq.~(\ref{equiv2}) in accordance with the principle of maximum entropy, and neglecting exponentially small corrections $\sim \mathrm{e}^{S_2-S_1}$. \\
In the approach of Schwinger, the focus of the SPA analysis was to provide a way to justify the ensemble equivalence as follows. Following a similar procedure as the one leading to Eq.~(\ref{density}), we write down the microcanonical density matrix $\hat{\rho}_{{\cal AB}}$ in Eq.~(\ref{micro1}) as 
\begin{equation}
\begin{aligned}                 
\hat{\rho}_{{\cal AB}}(E) 
&= \frac{1}{2\pi \mathcal{G}_{{\cal AB}}(E)} \\
\times &\int_{-\infty}^{\infty} d\tau \, \mathrm{e}^{i\tau E} 
\mathrm{e}^{\log Z_{{\cal AB}}(i\tau)}  \frac{\mathrm{e}^{-i \tau \hat{H}_{{\cal AB}}}}{Z_{{\cal AB}}(i\tau)},
\end{aligned}
\end{equation}
where we have multiplied and divided by $Z_{{\cal AB}}(i\tau)$. The expectation value of an observable $\hat{O}$ acting on the global system can then be written in the form
\begin{equation}\label{equiv}
%\begin{aligned}                 
\braket{\hat{O}}_E 
= \frac{1}{2\pi \mathcal{G}_{{\cal AB}}(E)} \\
\int_{-\infty}^{\infty} d\tau \, \mathrm{e}^{i\tau E} 
\mathrm{e}^{\log Z_{{\cal AB}}(i\tau)}\braket{\hat{O}}_{i\tau},
%\end{aligned}
\end{equation} 
where $\braket{\hat{O}}_{i\tau}=\Tr_{{\cal AB}}[\hat{O} \mathrm{e}^{-i \tau \hat{H}_{{\cal AB}}}/Z_{{\cal AB}}(i\tau)]$. In the thermodynamic limit, provided $\braket{\hat{O}}_{i\tau}$ varies slowly with respect to $\tau$ \cite{libroQSFT}, the integral in Eq.~(\ref{equiv}) can also be solved by SPA, resulting in $\braket{\hat{O}}_{E} \approx  \braket{\hat{O}}_{\beta(E)} $, where $\beta(E)$ is the solution of Eq.~(\ref{condition}). This is the meaning of equivalence of ensembles, according to Schwinger, in the weak coupling approach. Note that Eq.~(\ref{equiv}) is a mathematical identity which relates the expectation value of a global observable $\hat{O}$ calculated in the microcanonical equilibrium with the quantity $\braket{\hat{O}}_{i\tau}$, which in the thermodynamic limit will give the canonical expectation value of the observable evaluated at the inverse temperature $\beta(E)$. This identity then does not refer to any dynamical process of equilibration in time. The topic of dynamical equilibration or relaxation goes beyond the formalism developed in this paper.
In summary, Eq.~(\ref{equi}) establishes a relation of energy equilibrium between two many-body systems. The condition of equilibrium is fixed by the inverse temperature $\beta$ coming from the SPA analysis of Eq.~(\ref{SPA}) in the thermodynamic limit. We establish this limit considering, for example, a many-body global system $\cal A+\cal B$ with a constant energy per particle, where the total energy scales with the number of particles $N$. In that case the thermodynamic limit $N\to \infty$ gives the relation Eq.~(\ref{equiv2}). It is satisfactory to see how the SPA analysis formalizes the difference between the scenario of mutual equilibrium, where both subsystems are macroscopic and the temperature emerges from the distribution of the total energy where both systems get a finite fraction, and the bath scenario where $\nu_{{\cal A}}=0$ in Eq.~(\ref{phase2}) and the temperature of the subsystem is simply inherited from the temperature of the bath. In this last scenario the function ${\rm e}^{\log\, Z_{{\cal A}}(i\tau)}=Z_{{\cal A}}(i\tau)$ is smooth and does not participate of the SPA condition.

\section{Finite coupling regime}
After this revision of the key aspects of the emergence of temperature in composite weakly interacting systems, we now proceed to extend these ideas to systems with non-negligible interaction Hamiltonian $\hat{H}_{\text{int}}$. \newline
The key point that allow for this generalization is that the SPA analysis that naturally leads to the concept of temperature is not restricted to $\gamma \to 0$ at all, and in fact its only requirement is consistency with the thermodynamic limit $N\to \infty$.
While the global equilibrium state in the case of finite interaction energy is 
\begin{equation}\label{equi2}
  \hat{\rho}_{{\cal AB}}=\frac{\delta(E-\hat{H}_{{\cal A}}-\hat{H}_{{\cal B}}-\hat{H}_{\text{int}})}{\mathcal{G}_{{\cal AB}}(E)},
\end{equation}  
and the density of states is still given by Eq.~(\ref{density}), now $ Z_{{\cal AB}}(i\tau)$ can not be unambiguously decomposed in general in terms of the bare Hamiltonians $\hat{H}_{{\cal A}}$ and $\hat{H}_{{\cal B}}$ \cite{spheat}. However, our key observation is that {\it as long as we can solve the integral in Eq.~(\ref{density}) by SPA, the resulting real solution for $\beta$, which now depends not only on the total energy $E$ but also on the parameters of the interaction, characterizes the condition of thermal equilibrium between ${{\cal A}}$ and ${{\cal B}}$, thus providing the statistical definition of temperature for systems with finite coupling}. To support this claim we will now study the consistency and consequences of this definition in a solvable example. \\

As a specific microscopic model that allows for almost full analytical treatment and remains of high experimental relevance, we consider now a microcanonical modification of the widely used open-system approach to Quantum Brownian Motion (QBM)\cite{Ingold}. Here ${{\cal A}}$ consists of a quantum harmonic oscillator linearly coupled to a bath ${{\cal B}}$ of $N$ non-interacting harmonic oscillators.  The total Hamiltonian reads
\begin{equation}\label{hamilt}
\begin{aligned}
 \hat{H}_{\mathcal{A}\mathcal{B}}=&\frac{\hat{p}^{2}}{2m}+\frac{1}{2}m\omega_{0}^{2}\hat{q}^{2}\\ &+\frac{1}{2}\sum_{n=1}^{N}\Bigg[\frac{\hat{p}_{n}^{2}}{m_n}+m_n\omega_{n}^{2}\bigg(\hat{q}_{n}-\frac{c_n}{m_n\omega_{n}^{2}}\hat{q}\bigg)^{2}\Bigg],
\end{aligned} 
\end{equation}
where $\hat{p}$ and $\hat{q}$ are momentum and position operators of the coupled harmonic oscillator with bare frequency $\omega_0$ and mass $m$, and $\hat{p}_n, \hat{q}_n$ the momentum and position operators of the $n$th bath oscillator with frequency $\omega_n$ and mass $m_n$ coupled with the central system through characteristic $c_n$ fixed for a given model of the bath. \newline 
The bath and interactions are characterized by the bare and coupled spectral densities that distribute the frequencies $\omega_n$ like 
\begin{subequations}
\begin{eqnarray} 
\label{spectral}
I(\omega)&=&\pi \sum_{n=1}^{N}\delta(\omega-\omega_{n})=\kappa\omega^{2}e^{-\omega/\omega_{D}} \\ 
\label{spectral1}
J(\omega)&=&\pi \sum_{n=1}^{N}\frac{c_n^2}{2m\omega_n}\delta(\omega-\omega_n)=m\gamma \omega e^{-\omega/\omega_D}, 
\end{eqnarray}     
\end{subequations}
with cut-off Drude frequency $\omega_D$ and so called damping parameter $\gamma$, which is a function of the parameters $c_n^2$ characterizing the system-bath coupling strength. The parameter $\kappa$ is a characteristic of the bath with units of $\omega^{-3}$ such that $\int_{0}^{\infty} d\omega I(\omega)/\pi =N$.

In order to use Eq.~(\ref{density}), we construct $Z_{{\cal AB}}(i\tau)$ of the QBM model by analytical continuation of the Matsubara frequencies $\nu_{n}=\frac{2\pi n}{\hbar i\tau}$ from the known result \cite{Weiss}, to get
\begin{equation} \label{partition}
\begin{aligned}
 Z_{{\cal AB}}(i\tau)&=Z_{{\cal B}}(i\tau) \\ &\times \frac{1}{\hbar i\tau\omega_{0}}\prod_{n=1}^{\infty}\frac{\nu_{n}^{2}(\omega_{D}+\nu_{n})}{(\omega_{0}^{2}+\nu_{n}^{2})(\omega_{D}+\nu_{n})+\nu_{n}\gamma\omega_{D}}.
\end{aligned} 
\end{equation}
Here the imaginary temperature partition function $Z_{{\cal B}}(i\tau)$ of ${{\cal B}}$, using the spectral density from Eq.~(\ref{spectral}), reads 
\begin{equation} \label{logZb}
 \log Z_{{\cal B}}(i\tau)=-i\tau E_{0}+\frac{ 2 \kappa \zeta(4)}{(\hbar i\tau)^{3}}, 
\end{equation}  
where $E_0=3\kappa \hbar \omega_D^4$ is the zero point energy of the bath, and $\zeta(x)$ is the Riemann zeta function. In this way we arrive at
\begin{equation} \label{logZ}
\log Z_{{\cal AB}}(i\tau)=\log Z_{{\cal B}}(i\tau) + \log \tilde{Z}(i\tau),
\end{equation} 
where the effective $\tilde{Z}$, related to the coupled harmonic oscillator, has an explicit form in terms of Gamma functions $\Gamma$ \cite{Weiss},
\begin{equation}\label{gamma}
 \tilde{Z}(i\tau)=\frac{\hbar i\tau\omega_{0}\Gamma(\hbar i\tau\lambda_{1}/2\pi)\Gamma(\hbar i\tau\lambda_{2}/2\pi)\Gamma(\hbar i\tau\lambda_{3}/2\pi)}{4\pi^{2}\Gamma(\hbar i\tau\omega_{D}/2\pi)},
\end{equation}
with $\lambda_1, \lambda_2$ and $\lambda_3$ being the roots of the polynomial expression in $\nu_{n}$ that appears in the denominator of Eq.~(\ref{partition}) and which carry the dependence on $\gamma, \omega_D$ and $\omega_0$. \\

Interestingly, as shown in \cite{pachon}, for systems that have an interaction that involves only relative coordinates, like in Eq.~(\ref{hamilt}), the \textit{classical} partition function does not depend at all on the coupling 
\begin{align*}
\log Z^{\text{classic}}_{{\cal AB}}(i\tau)=\log Z^{\text{classic}}_{{\cal B}}(i\tau) + \log Z^{\text{classic}}_{{\cal A}}(i\tau),
\end{align*}
and therefore the temperature $\beta$ is independent of $\gamma$, regardless how strong the interaction is. This means that for the model in Eq.~(\ref{hamilt}) \textit{the dependence of the temperature on the coupling strength is purely a quantum effect}. \\

Before going further we make an important remark. Since our model consists of a single harmonic oscillator (system $\cal A$) linearly coupled to many harmonic oscillators (system $\cal B$), we find here the situation where the SPA analysis requires $\nu_{{\cal A}}=0$ as discussed in the last part of section \ref{section1}. Since the function $\tilde{Z}$ in Eq.~(\ref{logZ}) encloses the effect of {\it both} the single harmonic oscillator plus interaction terms, we must subtract from Eq.~(\ref{logZ}) the term $\log Z_{\cal A}(i\tau)$ due to the bare central oscillator, meaning that we are considering $Z_{\cal A}(i\tau)$ smooth enough not to let it participate in the SPA analysis of Eq.~(\ref{density}). With this in mind we now use Eqns.~(\ref{density}, \ref{logZb}, \ref{logZ}) to identify the action
\begin{equation}\label{action2}
\phi(i\tau)=i (E-E_0)\tau +\frac{ 2 \kappa \zeta(4)}{(\hbar i\tau)^{3}} +\log \tilde{Z}(i\tau)-\log Z_{\cal A}(i\tau).
\end{equation} 

Solving the integral in Eq.~(\ref{density}) by SPA, using the saddle-point condition in Eq.~(\ref{SPA}),
and again, looking for real solutions $\beta=i\tau^{*}$, we get
\begin{equation}\small  \label{res1}
\begin{aligned}
 E-E_0-\frac{ 6 \kappa \zeta(4)}{\hbar^{3}}\frac{1}{\beta^{4}}  = \frac{1}{\beta} -\frac{\hbar \omega_0}{2}\coth\Big(\frac{\beta \hbar \omega_0}{2}\Big) \\ +\frac{\hbar}{2\pi}\Bigg\{\omega_{D}\psi(1+\hbar\beta\omega_{D}/2\pi)-\sum_{i=1}^{3}\lambda_{i}\psi(1+\hbar\beta\lambda_{i}/2\pi)\Bigg\},
\end{aligned}
\end{equation}
where we have inserted Eq.~(\ref{gamma}) into (\ref{action2}). Note the subtracted energy of the bare oscillator $\frac{\hbar \omega_0}{2}\coth\Big(\frac{\beta \hbar \omega_0}{2}\Big)$. Here $\psi(x)=\frac{d}{dx}\log\Gamma(x)$ denotes the Digamma function. \\

Eq.~(\ref{res1}) establishes the equilibrium relation for the total energy $E$ between system ${{\cal B}}$ and the interaction energy, where the \textit{l.h.s} is related with the energy of the bath and the \textit{r.h.s} accounts for the energy of interaction. \newline 
In the regime where $\gamma=0$ the \textit{r.h.s} of Eq.~(\ref{res1}) is zero. In this case the derived temperature is given by the bath. This is the common scenario in the weak-coupling canonical approach, where the central system acquires a temperature given by the constant temperature of the canonical thermal bath. Here we will show that an interaction term that couples linearly the system with each degree of freedom of the bath gives rise to an interaction energy which affects the resulting equilibrium temperature.   \\

 The divergences affecting Eq.~(\ref{res1}) for $\omega_D  \to \infty$ arise from the well known \cite{renorm} divergences of the ground state energy of the coupled harmonic oscillator $\epsilon_0$, 
\begin{equation*}
\begin{gathered}
\epsilon_0=\frac{\hbar}{2\pi} \Bigg[\lambda_1 \log[\omega_D/\lambda_1]+\lambda_2 \log[\omega_D/\lambda_2] + \lambda_3 \log[\omega_D/\lambda_3]\Bigg],
\end{gathered}
\end{equation*} 
but are readily renormalized by $\tilde{Z}\times e^{\beta \epsilon_0}$ to obtain a global zero ground state energy. The new relation (\ref{res1}) for renormalized energy finally reads
\small
 \begin{equation} \label{res2}
 \begin{aligned}
 E -\frac{ 6 \kappa \zeta(4)}{\hbar^{3}}\frac{1}{\beta^{4}}  & =  \frac{1}{\beta}+\frac{\hbar}{2\pi}\Bigg\{\omega_{D}\psi(1+\hbar\beta\omega_{D}/2\pi)-\omega_D\log \omega_D\Bigg\}   \\ &-\frac{\hbar}{2\pi}\sum_{i=1}^{3}\Bigg(\lambda_{i}\psi(1+\hbar\beta\lambda_{i}/2\pi)-\lambda_i\log \lambda_i\Bigg)\\& -\frac{\hbar \omega_0}{2}\coth\Big(\frac{\beta \hbar \omega_0}{2}\Big)+\frac{\hbar \omega_0}{2}, 
\end{aligned}
\end{equation} 
\normalsize 
where we have made use of the Vieta relation $\lambda_1+\lambda_2+\lambda_3=\omega_D$ \cite{Weiss}.  
The solution of Eq.~(\ref{res2}) for $\beta$ (fixing the energy equilibrium condition for our model) accordingly defines the inverse temperature in the finite coupling regime. \newline 
This solution $\beta(E,\gamma)$, our main result, depends on the interaction $\gamma$, the total energy $E$, but also the bath parameters $\kappa$ and $\omega_D$. \newline
Coming back to the issue of SPA vs weak coupling expansions, we stress again that in a model where the total energy $E$ scales with the number of particles $N$, in solving Eq.~(\ref{density}) by SPA we are neglecting terms of $\mathcal{O}(1/\sqrt N)$, which is justified in the thermodynamic limit for large number of particles \cite{libroQSFT}. Still, the SPA approximation does not depend on any perturbative expansion of the interaction parameter $\gamma$ and thus our results are valid beyond the weak-coupling limit. \\
  
\begin{figure}[]
%\centering
\includegraphics[width=0.4\textwidth]{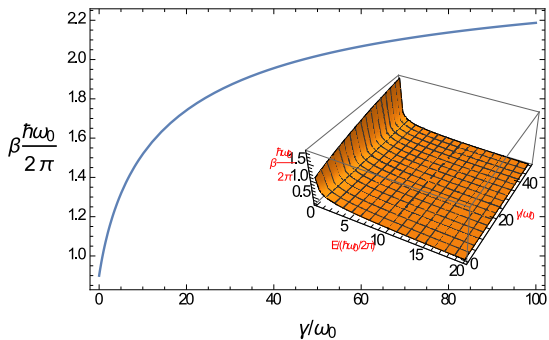}
\caption{Inverse temperature $\beta(\gamma)$ for given values $\kappa \omega_0^3=5$, $\omega_D/\omega_0=10$, and $E/\Big( \frac{\hbar \omega_0}{2\pi}\Big)=0.2$, showing the increase of $\beta$ with $\gamma$ near the ground state energy. Inset shows the variation of $\beta(E,\gamma)$ for a large range of energies, showing a monotonic decrease with $E$. }
\label{Fig1}
\end{figure}

\begin{figure}[]
	\includegraphics[width=0.4\textwidth]{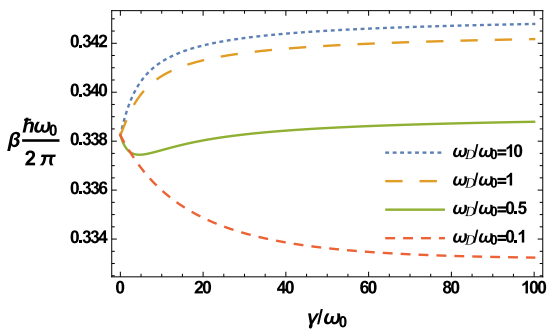}
	\caption{$\beta(\gamma)$ for parameters $\kappa \omega_0^3=5$ and $E/\Big( \frac{\hbar \omega_0}{2\pi}\Big)=10$, showing the increasing of $\beta$ in the Markovian regime and its change of behavior in the non-Markovian case.}
	\label{Fig2}
\end{figure}

\begin{figure}
%	\centering
	\includegraphics[width=0.4\textwidth]{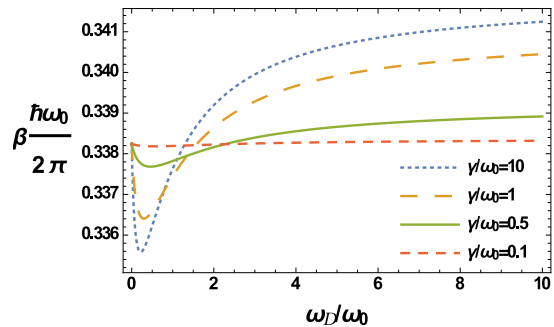}
	\caption{$\beta(\omega_D)$ for parameters $\kappa \omega_0^3=5$ and $E/\Big( \frac{\hbar \omega_0}{2\pi}\Big)=10$, showing the explicit dependence of $\beta$ with the Drude frequency and its change of behavior from Markovian to non-Markovian regime.}
	\label{Fig3}
\end{figure}

In Fig.~\ref{Fig1} we show the numerical solution $\beta(E,\gamma)$ of Eq.~(\ref{res2}) for given values of $\kappa$ and $\omega_D$ and for energy near the renormalized ground state where quantum effects are more visible. A clear variation of the temperature as a function of $\gamma$ is observed, showing that the interaction energy has a sensible effect in the derived temperature. In the inset of Fig.~\ref{Fig1}  
 can be observed that $\beta$ is a monotonous function of the total energy $E$. From Eq.~(\ref{res2}), and using the asymptotic expansion of Digamma functions, we obtain that for $\beta \to \infty$, $E \to 0$, as expected. For large energies the change of $\beta$ with $\gamma$ is less evident, but still can be appreciated in Fig.~\ref{Fig2}. Remarkably in this regime the behavior of $\beta$ with $\gamma$ is modulated by the tunning parameter $\omega_{D}/\omega_{0}$, that also quantifies the degree of memory of the bath or non-Markovianity \cite{paz}. This surprising connection between the dependence of the temperature on the coupling and the time scale $2\pi/\omega_{D}$  associated to memory effects in the environment is also shown by the explicit dependence of $\beta$ on the Drude frequency in Fig.~\ref{Fig3}. \\

We can provide a physical interpretation of the peculiar dependence of $d\beta / d \gamma$ on the Drude frequency $\omega_{D}$ for our results in the following way: 
In the \textit{Markovian regime} Fig.~\ref{Fig2} suggest that the action from the bath on the central system mostly determines the energy equilibrium condition: the energy-flow follows the natural direction from bath to system to reach equilibrium. Moreover, it can be shown that the \textit{r.h.s} of (\ref{res2}) grows with $\gamma$ in this regime. That explains the growing behavior of $\beta$ with $\gamma$ in this case. On the other hand, in the \textit{non-Markovian regime} Fig.~\ref{Fig2} suggest that is the action from the system on the bath which determines the equilibrium condition: in this case the energy-flow direction goes from system to bath.  Also, it can be shown that the contribution of the \textit{r.h.s} of (\ref{res2}) in this case follows a decreasing or increasing behavior as a function of $\gamma$ depending on the particular range of values of $\omega_D/\omega_0$ within the non-Markovian regime. This is reflected in the peculiar behavior appearing in Fig.~\ref{Fig2}.    \\ 

\begin{figure}
	\centering
	\includegraphics[width=0.4\textwidth]{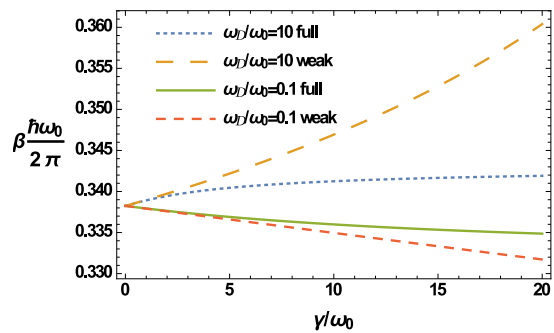}
	\caption{Contrast between the temperature solution at first order expansion in $\gamma$  and the full results, showing the characteristic saturation behavior of the full result in contrast with the first order perturbative expansion in $\gamma$.}
	\label{Fig_neu}
\end{figure}

We want to emphasize that the results obtained here go beyond any finite-order expansion around the weak-coupling scenario. Fig.~\ref{Fig_neu} shows the contrast between the solution for temperature obtained in the first order expansion for $\gamma$ in Eq.~(\ref{res2}) and that obtained from the full expression. The characteristic saturation behavior clearly indicates the breakdown of any finite-order approximation in powers of $\gamma$. \\

\begin{figure}
	\centering
	\includegraphics[width=0.5\textwidth]{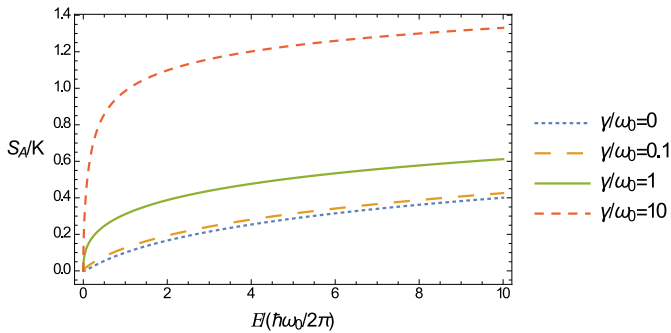}
	\caption{Subsystem entropy $S_{{\cal A}}(E,\gamma)$ for parameters $\kappa \omega_0^3=5$ and $\omega_D/\omega_0=10$. The entropy of the subsystem $\cal{A}$ is parametrically larger when the system-depended damping $\gamma$ is large and also increases with the energy, and becomes zero for $E=0$, in accordance with the laws of thermodynamics. }
	\label{Fig4}
\end{figure}

Having obtained the inverse temperature $\beta(E,\gamma)$ we can calculate thermodynamic potentials for finite coupling. We start with Eq.~(\ref{equiv2}), from which the entropy of the {\it global} system in the limit $N \to \infty$ can be calculated by
\begin{equation}
\log Z_{{\cal AB}}(\beta)=S(E)/K-\beta E.
\end{equation}
Recognizing that $\beta(E,\gamma)$ is a function of $E$, we get
\begin{equation}
\frac{1}{K}\frac{\partial S}{\partial E}=\beta +\frac{\partial \beta}{\partial E}\Bigg(E+\frac{\partial}{\partial \beta}\log Z_{{\cal AB}}\Bigg),
\end{equation} 
and since $\frac{\partial}{\partial \beta}\log Z_{{\cal AB}}=-E$, we finally obtain the thermodynamic relation
\begin{equation}
\frac{1}{K}\frac{\partial S}{\partial E}=\beta(E,\gamma).
\end{equation}
Following Ref. \cite{entropy3} we may also calculate the entropy for the coupled oscillator as
\begin{equation}
\frac{S_{{\cal A}}}{K}=\log \tilde{Z}(\beta)-\beta \frac{\partial}{\partial \beta}\log \tilde{Z}(\beta),
\end{equation}
where $\tilde{Z}$ is taken from Eq.~(\ref{gamma}) and evaluated at the solution $\beta(E,\gamma)$ given by the SPA condition. The term $-\frac{\partial}{\partial \beta}\log \tilde{Z}(\beta)$ is the thermodynamic mean energy of the coupled oscillator evaluated at $\beta(E, \gamma)$. \\
Fig.~\ref{Fig4} illustrates the behavior of $S_{{\cal A}}$ as a function of the total energy $E$ for various values of the coupling $\gamma$. As can be observed, $S_{{\cal A}}$ is a positive quantity that becomes zero for $E=0$, in nice accordance with the third law of thermodynamics. The entropy is also a monotonically increasing function of $E$ and $\gamma$. This latter feature accounts for the decrease in purity of the reduced density matrix ${\rm Tr}_{{\cal B}}~\hat{\rho}_{{\cal A} {\cal B}}$ with increasing $\gamma$. \\ 

Finally, let us consider the finite coupling version of the ensemble equivalence. Whenever the solution of the integral in Eq.~(\ref{equiv}) is justified by SPA, the saddle-point condition will give a relation between the expectation value of any smooth operator calculated in the microcanonical ensemble and the one evaluated in the canonical case, \textit{but for a temperature given by the solution $\beta(E,\gamma)$ in the finite coupling regime}. The same considerations also hold for the reduced density matrix describing the subsystem ${{\cal A}}$ and, in that case, the relation for the expectation value of an observable $\hat{O}_{{\cal A}}$ is given as
\begin{equation}
\braket{\hat{O}_{{\cal A}}}_E \approx \braket{\hat{O}_{{\cal A}}}_{\beta(E,\gamma)},
\end{equation}   
\begin{figure}
%	\centering
	\includegraphics[width=0.4\textwidth]{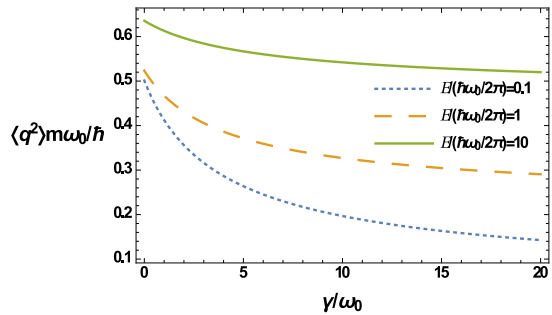}
	\caption{$\braket{q^2}(\gamma)$ for parameters $\kappa \omega_0^3=5$ and $\omega_D/\omega_0=10$. As expected the particle gets more localized with increasing $\gamma$ and its squared position expectation value increases with $E$.}
	\label{Fig5}
\end{figure} 
that provides the sought extension of the equivalence of ensembles for systems with finite coupling in the thermodynamic limit. Accordingly, in Fig.~\ref{Fig5} we show the expectation value of the squared position operator of the coupled oscillator evaluated at the solution $\beta(E,\gamma)$. As expected $\braket{q^2}$ grows with the energy $E$ and the particle is getting more localized with the increase of the damping parameter $\gamma$, as the bath monitors the position of the central particle \cite{carlos}. \\
It is interesting to note that a similar behavior of the entropy and the expectation value of the squared position with respect to $\gamma$ in the QBM model is found in the canonical thermal bath approach \cite{entropy3,Weiss}. This is actually a non-trivial result due to the fact that our microcanonical thermodynamics is based on $\beta(E,\gamma)$, which is a SPA condition solution that involves $\gamma$. One could imagine a situation where an observable in Eq.~(\ref{equiv}) has a non-smooth dependence with the integration variable $\tau$, and therefore this dependence must be included in the SPA condition. In such scenario the equilibrium temperature becomes itself a function of the particular observable and our notion of \q{ensemble equivalence} must be accordingly modified.

\section{Conclusions}
 Starting from the fundamental microcanonical distribution we have studied the emergence of temperature $T$ in the finite coupling regime of open quantum systems. Following the approach pioneered by Schwinger resulting in $T=T(E)$ as an emergent quantity which establishes the condition of equilibrium between two weakly coupled subsystems at energy $E$, we have shown that in the finite and strong coupling regime $T=T(E,\gamma)$ also depends on the parameter $\gamma$ that characterize the strength of the interaction. We have applied this idea to the paradigmatic Quantum Brownian Motion model and studied the main features of this notion of temperature, confirming that $T(E,\gamma)$ is a monotonically increasing function of the total energy $E$, and showing a clear variation of $T$ with $\gamma$ which is  a purely quantum effect particularly visible near the ground state energy. The entropy of the coupled oscillator, which now depends on $\gamma$, is a positive quantity that starts from zero for $E=0$ and increases monotonically with $E$ and $\gamma$. Remarkably, we found also, for large energies, an unexpected dependence on the memory properties of the bath: while $T(E,\gamma)$ decreases as a function of the interaction parameter $\gamma$ in the Markovian regime, the behavior is different in the non-Markovian case. \\
\section{Outlook}
 In a future work we want to extend this formalism to the cases where the central system is composed of non-interacting bosons or fermions which, besides energy, can also interchange particles with a reservoir. When a microscopic model is identified one must first extend the present formalism to the many-body context including correct fermion/boson statistics. Then, one can obtain the change of the temperature with respect to the coupling parameter, in the spirit of Fig.~\ref{Fig2}, and see how much the value of the temperature deviates from the temperature of the bath alone, a deviation that can be measured. In this case our formalism could be connected to recent experiments where thermodynamic quantities of few particles has been measured \cite{experiment}. This, and the effect of finite coupling in the degeneracy of the ground state of the isolated system \cite{degeneracy1, degeneracy2} remain interesting questions for further study.   
\\
 
\textbf{Acknowledgments}. C.A.M. acknowledges financial support from the German Academic Exchange Service DAAD. 
We want to thank Josef Rammensee for useful discussions, and an anonymous referee for interesting suggestions.

% General remaks: 1. We claim that whenerver we can solve the density of states by SPA the tmeperature emerges. But we cannot say aniything about what to do if SPA is not justified. In particular, for small size global systems, or systems that has a energy bound from above, which densisty of states doestn grow exponentially with energy, we dont know how to define a temperature in the strong coupling regime in these cases. 
% 2. Our central resutl can be interpreted in terms of the zeroth law of thermodynamcis for strong coupled. (What about the trnasitiviyt of the thermal equilibrium ?)
% Referies: Prof. Dr. Gert-Ludwig Ingold. Institut für Physik. Universität Augsburg D-86135 Augsburg, Germany.  Gert.Ingold@Physik.Uni-Augsburg.de
%  Dr. Eric Lutz Institut für Physik. Universität Augsburg D-86135 Augsburg, Germany,  	Eric.Lutz@physik.uni-augsburg.de
% Dr. Jens Siewert. Departamento de Qu ́ımica F ́ısica, Universidad del Pa ́ıs Vasco UPV/EHU, E-48080 Bilbao, Spain.  jens.siewert@physik.uni-regensburg.de 
%

\begin{thebibliography}{56}
\bibitem{pie}
The operator-valued function $\delta(E-\hat{H})$ is defined as: $\delta(E-\hat{H})=\sum_{n}\ket{E_n}\bra{E_n}\delta(E-E_n)$, in the energy eigenbasis of $\hat{H}$.
\end{thebibliography}
\end{document}